\documentclass{aa}  
\newdimen\imgwidth
\usepackage{graphicx}
\usepackage[usenames]{color}
\usepackage[labelsep=period,labelfont=bf]{caption}
\usepackage{amsfonts}
\usepackage{amsmath}
\usepackage{txfonts}

\begin{document}
   \title{The environment of AGNs and the activity degree of their surrounding
          galaxies 
\thanks{Based on Sloan Data}
 }

   \author{W. Kollatschny \inst{1},
           A. Reichstein \inst{1,2},
          M. Zetzl \inst{1} 
          }

   \institute{Institut f\"ur Astrophysik, Universit\"at G\"ottingen,
              Friedrich-Hund Platz 1, D-37077 G\"ottingen, Germany\\
              \email{wkollat@astro.physik.uni-goettingen.de}
           \and  
              University College Cork, Ireland\\ 
}


   \date{Received 2011; accepted 2012}
   \authorrunning{Kollatschny, Reichstein, Zetzl}
   \titlerunning{Environment of AGNs}


 \abstract{}
{We present results of a comprehensive spectral study on the
large-scale environment
of AGNs based on Sloan Spectroscopic Survey data.} 
{We analyzed the spectra of galaxies in the environment of AGN and other
 activity classes up to distances of 1 Mpc.}
{The mean H$\alpha$ and [\ion{O}{iii}]\,$\lambda 5007$ line luminosities
in the environmental galaxies within a projected radius of 1~Mpc are highest
around Seyfert~1
galaxies, with decreasing luminosities for Seyfert~2 and HII galaxies,
and lowest for absorption line galaxies.
Furthermore, there is a trend toward H$\alpha$ and [OIII] luminosities
in the environmental galaxies increasing as a function of proximity to
the central emission line galaxies. There is another clear trend toward a
neighborhood effect within a radius of
1000 kpc for the AGN and non-AGN types: Seyfert galaxies tend to have
the highest probability of having another Seyfert galaxy in the
 neighborhood. HII galaxies tend to have
the highest probability of having another HII galaxy in the
 neighborhood, etc.
The number of companions within 1000 kpc is inversely correlated with
the  H$\alpha$, [\ion{O}{iii}]\,$\lambda 5007$, as well as
with the continuum luminosities of the central galaxies,
regardless of whether they are of Seyfert, HII, or absorption
line types.
}
{}

\keywords {Galaxies: active --
                Galaxies: Seyfert  --
                Galaxies: starbursts  --
                Galaxies: interactions --
                Galaxies: statistics
               }

   \maketitle
%

\section{Introduction}
It has been known for many decades that the morphology of galaxies is correlated
to the number of galaxies in their environment. For instance, Oemler
(\cite{oemler77}) noted that the population of
elliptical, spiral, and S0 galaxies
varies with the mean density of galaxy clusters. This so-called
morphology-density relation (MDR) might be explained by the slow stripping of
gas from spiral galaxies (e.g. van der Wel et al. \cite{vanderwel10}).

In addition to this finding, astronomers have been debating
since more than 20 years whether the environment
of active galactic nuclei (AGN) or starburst galaxies is different
from that of inactive
galaxies. This idea is based on the concept that nuclear
activity, as well as starburst activity in galaxies, is
triggered by galaxy interactions and/or that in general the environment
 of active/non-active galaxies
is different for different types
(e.g. early papers by Hutchings et al. \cite{hutchings84},
Dahari \cite{dahari85}, Keel et al. \cite{keel85},  Bushouse \cite{bushouse90}, 
Byrd et al. \cite{byrd87}).

In the meantime it has been accepted by most astronomers that starburst
activity in galaxies is triggered by tidal interactions
(e.g. Woods \& Geller \cite{woods07}, Shardha et al., \cite{shardha09} and
references therein).
However, the concept of tidal triggering for the generation of nuclear
Seyfert/Quasar activity is less clear.
Many studies have been undertaken to investigate the environmental density of
active galaxies and to compare it with inactive and/or starburst galaxies.
Some authors claim an underdensity of
bright galaxies in the environment of Seyfert galaxies/quasars
(e.g. Constantin \& Vogeley, \cite{constantin06}; Lietzen et al.,
\cite{lietzen09}),
while others find a similar clustering strength in the environment
of AGN and inactive galaxies
on scales of less than 100 kpc up to scales larger than a few Mpc
(e.g. Li et al., \cite{li06}).

Other studies have investigated the influence of large-scale structures on the
color distribution of AGN host galaxies. 
Silverman et al. (\cite{silverman08}) demonstrate that the
color distribution of AGN
host galaxies is highly dependent on the influence of ~10 Mpc scale structures.
Coldwell et al.
(\cite{coldwell09}) report that the neighborhood of their blue AGN sample is
indistinguishable from their inactive counterpart, while the
red AGN environments show an excess of blue star-forming galaxies.

 Furthermore, it is under debate whether the environment 
is different for different types of AGN. 
Based on the unified model of AGN one would expect no difference in the
environment of Seyfert 1 and Seyfert 2 galaxies. Corresponding to this picture,
Sorrentino et al. (\cite{sorrentino06}) extracted galaxies in the
redshift range $0.05 \le z \le 0.095$ from the Fourth Data Release (DR4) of the
Sloan Digital Sky Survey (SDSS).
They see no difference in the large-scale
environment of the two Seyfert types.
Koulouridis et al. (\cite{koulouridis06})
and Strand et al. (\cite{strand08})
investigated the environment of Seyfert galaxies
having redshifts of
$0.004 \le z \le 0.036$
and $0.11\leq z\leq 0.6$, respectively.
They, however, find an overdensity of companions around
Seyfert 2 galaxies.

Nearly nothing is known about the spectral activity of galaxies
in the environment of AGN
(Kollatschny \& Fricke, \cite{kollatschny89}, Fricke \& Kollatschny,
\cite{fricke89}). 
Here we carry out an in-depth analysis of the  spectral activity of the
 environmental galaxies of 3500 AGN/non-AGN
 based on Sloan Digital Sky Survey (SDSS) spectra.
We investigate the spectral properties of the  environmental galaxies
up to projected distances of 1 Mpc.

Throughout this paper we use
$H_{0} = 75 km\,s^{-1}\,Mpc^{-1}$.
We neglect cosmological corrections on the distances of our
galaxies as they are local (z$\leq$0.08).

The paper is organized as follows.
The data samples and the algorithms we used to find the number of neighbors
 are described in Sects. 2 to 4.
The results regarding the number of companions and the activity of the
environmental galaxies are presented in Sect. 5.
A detailed discussion and the conclusion are given in Sect. 6.

\section{The SDSS spectroscopic survey}

The present environmental study is based on spectra from the Sloan
Digital Sky Survey (SDSS) Data Release 5 (DR5; Adelman-McCarthy
(\cite{adelman07}). The SDSS is a photometric and spectroscopic survey of a
quarter of the sky. The SDSS DR5 covers an area of 5740 square degrees. 
Imaging and positional data are available in the u,g,r,i,z bands
for more than 100 million objects.
The imaging data are calibrated photometrically and astrometrically.

More than one million spectra are contained in the SDSS DR5.
The spectra cover the spectral range 3800 $\le \lambda \le$ 9200 \AA\ with a
spectral resolution of $\lambda / \delta\lambda$\,=\,1800. This corresponds
to a redshift accuracy of $\sim170\,km\,s^{-1}$.

\section{Definitions of AGN/non-AGN samples}

The spectra of our current study were selected from the spectroscopic
SDSS survey. First we selected all objects that have been classified as galaxies
or quasars. In a second step we restricted ourselves to nearby galaxies/QSOs
having redshifts z of less
than 0.08 to narrow down the sample. Intrinsically
faint environmental galaxies can be studied in detail only in the
nearby universe.
For detecting faint galaxy companions we limited
our sample to central AGN/non-AGN galaxies having g-band fiber magnitudes
of at least 18 (for fiber magnitude definitions see  Adelman-McCarthy
(\cite{adelman07}) and references therein).
This corresponds to g-band luminosities of e.g. L=3.2$\cdot 10^{41}$
erg\,s$^{-1}$ for z=0.01 
and L=7.4$\cdot 10^{42}$ erg\,s$^{-1}$ for z=0.05.
Galaxies in the nearby universe (e.g. M100) have a g-band luminosity of
L=1$\cdot 10^{43}$ erg\,s$^{-1}$.

In addition we set the condition that the given line fluxes of
 H$\beta$ and
[\ion{O}{iii}]\,$\lambda 5007$ be higher than
$10^{-17}\,erg\,s^{-1}\,cm^{-2}$  to exclude low signal-to-noise spectra from our
emission line sample.
 The absorption line galaxies
have negative H$\alpha$ and H$\beta$ line fluxes.

We generated two spectroscopic Seyfert samples (Seyfert~1, Seyfert~2 galaxies)
and two control samples (HII galaxies, absorption line galaxies) for
the present study of the AGN environment -- based on the SDSS
spectroscopic survey.

Seyfert~1 galaxies are identified on the basis of their broad Balmer emission
lines in the SDSS spectra.
The other active galaxy types were classified on the basis of their emission
line ratios of  [\ion{O}{iii}]\,$\lambda 5007$, H$\beta$,
[\ion{N}{ii}]\,$\lambda 6583$, and H$\alpha$ lines in the diagnostic diagram of
Kauffmann et al. (\cite{kauffmann03}).
We executed the following queries 
for selecting individual central galaxy types from the SDSS spectra:\\
\textbullet{} Seyfert~1 galaxies were by definition those galaxies having
 emission line widths (FWHM) broader than $1000\,km\,s^{-1}$.\\
\textbullet{} Seyfert~2 galaxies were selected on the
basis of the line intensity ratios of 
the emission lines [\ion{O}{iii}]\,$\lambda 5007$, H$\beta$,
[\ion{N}{ii}]\,$\lambda 6583$, and H$\alpha$.
They are defined by\\ 
$ \log{\left( [\text{O \sc iii}]\,\lambda 5007/\text{H}\beta\right)} > 0.61/ \{\log{ \left(  [\text{N\sc ii}] / \text{H}\alpha \right)} - 0.05 \} + 1.3$
and  ${[\text{NII}] / \text{H}\alpha}< 1.1220 $.\\
\textbullet{} HII galaxies are defined by\\
$ \log{\left( [\text{O \sc iii}]\,\lambda 5007/\text{H}\beta\right)} < 0.61/ \{\log{ \left(  [\text{N\sc ii}] / \text{H}\alpha \right)} - 0.05 \} + 1.3.$\\
\textbullet{} Absorption-line galaxies show
H$\alpha$ and H$\beta$ in absorption.

The galaxies in the AGN/non-AGN environment were divided by us
 into further subclasses
with respect to their emission/absorption line intensities:\\
 \textbullet{} The narrow-line AGN were divided
into Seyfert~2 and LINER types.
The dividing line between these two types 
is given by the subsequent line intensity ratios in the diagnostic diagram:\\
$ \log{\left( [\text{O \sc iii}]\,\lambda 5007/\text{H}\beta\right)} =$
$\log \left( \,{ [\text{N\sc ii}] / \text{H}\alpha } \,
\right)\, \cdot \tan(25\degr) \, + \,0.45\, \cdot \,\tan(25\degr)\, - \, 0.5 $\\
(Kauffmann et al., \cite{kauffmann03}; Kewley et al., \cite{kewley06}).\\
\textbullet{} $H\alpha$ emission line galaxies exhibit clear  $H\alpha$ emission lines.\\
\textbullet{} Absorption line type 1 galaxies (Abs1)
show Balmer lines in absorption with
weak additional emission lines,
such as
 $[\text{O \sc iii}]$ and $[\text{N\sc ii}]$ emission lines.\\
\textbullet{} Absorption line 2 galaxies (Abs2)
 display the Balmer lines in absorption, along with neither
  $[\text{O \sc iii}]$ nor $[\text{N\sc ii}]$ lines
in emission.

Based on the criteria given above
 we established four environmental samples around four different
types of central galaxies from the spectroscopic SDSS survey:
 two Seyfert samples consisting of 114 central Seyfert~1 and
1480 Seyfert~2 galaxies, as well as two control samples consisting of
1406 central HII galaxies and 415 absorption line galaxies (see Table 1).
The mean g-band continuum luminosities
(log $\bar{L}$ = 43.3 - 43.8 erg\,s$^{-1}$) 
of the central galaxies are equal within a factor of three in all samples with
$\bar{L}_{\text Sey1} = 4.3 \cdot 10^{43}$ erg\,s$^{-1}$ and
$\bar{L}_{\text Sey2} = 6.5 \cdot 10^{43}$ erg\,s$^{-1}$.

SDSS spectra exist only for a subsample of the SDSS photometric survey
because of the limited number of fibers, as well as because of the flux
limits for getting spectra with sufficient signal-to-noise ratio. 
There are typically five time more galaxies in the environmental regions
 - based on their colors -
in the photometric galaxy survey in comparison to the number
 of spectra that have been taken with SDSS.

\section{AGN environment and reference samples}

We inspected all SDSS spectra of the AGN/non-AGN
 companion galaxies within
a projected radius of 1000 kpc of the 3,415 central galaxies (Table\,1).
 The projected distance has been calculated by us by
means of the coordinates and the redshift of the
central galaxy.
 Furthermore, only those galaxies were selected as physical
neighbors to the central galaxy that have
redshifts within z = 0.003 of the central galaxy or
 have radial velocities within $\pm 800\,km\,s^{-1}$, respectively
(e.g. Madore et al., \cite{madore04}). 

In addition we inspected the SDSS direct images if 
 companion galaxies to other galaxies were 
 picked out by our SDSS DR5 query
within a projected distance of 15 kpc.
 We verified by eye on the direct images 
 whether the listed objects were really distinct galaxies.
In most cases the listed emission line spectra
were generated in HII regions within the host galaxies. 

All together we determined the number
of companion galaxies in the environment of the 3,415 central galaxies
(Seyfert~1, Seyfert~2, H\,II galaxies, absorption line galaxies) and
 analyzed the H$\alpha$ and [\ion{O}{iii}]\,$\lambda 5007$ line intensities
of 21,469 environmental galaxies.

\section{Results}

\subsection{The number of AGN/non-AGN galaxies and their companions}

In Table 1 we present the different activity types of central galaxies, their
numbers, the total amount of companions detected in a projected radius
of 1000 kpc, and the average number of companions per central galaxy. 
\begin{table}
\tabcolsep+2mm
\caption{Number of environmental galaxies within 1 Mpc.}
\centering
\begin{tabular}{c|r|r|r}
\hline 
\noalign{\smallskip}
Activity Type & Number of &  Number of & Average No. \\
of centr. gal.& centr. gal&  comp. gal.&  of comp. gal.\\
\hline 
\noalign{\smallskip}
 Sey1 & 114  &  514  & 4.51 \\
 Sey2 & 1,480 & 7,867  & 5.32 \\
 HII & 1,406 & 7,456  & 5.30 \\
 Abs & 415  & 5,632  & 13.57 \\
\hline
\noalign{\smallskip}
 & 3,415 & 21,469  & \\
\hline 
\end{tabular}
\end{table}
With an average number of 4.5, Seyfert~1 galaxies have the
 fewest companion galaxies within a projected distance of 1000 kpc.
Seyfert~2 Galaxies and
HII galaxies have about an equal number of companions with 5.3.  
 Absorption
line galaxies have with 13.6 by far the most companions.

In addition we counted the galaxies
within spherical shells around the central galaxies 
to test whether the relative number of companions 
is distinct for different types of central galaxies as a function of distance.
 The absolute number, as well as the average
number of companion galaxies (based on the numbers in Table 1),
 is given in Table 2 for different 
spherical shells.
\begin{table}
\tabcolsep+2mm
\caption{Number of environmental galaxies within 0 -- 0.2 Mpc,
as well as within spherical shells of 0.4 -- 0.6 Mpc, and
0.8 -- 1 Mpc.}
\centering
\begin{tabular}{c|r|c|r|c|r|c}
\hline 
\noalign{\smallskip}
Central& Comp.&avg No&Comp.&avg No&Comp.&avg No\\
   gal.&0--0.2&0--0.2&0.4--0.6&0.4--0.6&0.8--1&0.8--1.\\
\hline 
\noalign{\smallskip}
 Sey1 &   70 & 0.61 &  109 & 0.96 &  111 & 0.97 \\
 Sey2 & 1046 & 0.71 & 1659 & 1.12 & 1786 & 1.21 \\
 HII &  938 & 0.67 & 1559 & 1.11 & 1685 & 1.20 \\
 Abs &  612 & 1.47 & 1189 & 2.87 & 1077 & 2.60 \\
\noalign{\smallskip}
\hline 
\end{tabular}
\end{table}
The absolute number of companions is increasing in the shells as a
function of radius because the volumes are becoming systematically
bigger. 
However, the general difference regarding
 the average number of companions for the various
types of central galaxies
is distance independent.
\subsection{The number of companion galaxies as function of
the H$\alpha$, [\ion{O}{iii}]\,$\lambda 5007$, as well as
continuum luminosities of the central galaxies}

We identified the number of companion galaxies
as a function of the H$\alpha$, [\ion{O}{iii}]\,$\lambda 5007$,
as well as
g-band continuum luminosities of their central galaxies.
We determined this for all 21,469 galaxies within the
projected radius of 1 Mpc in the environment of the four different
types of central galaxies (Seyfert~1, Seyfert~2, HII, absorption line gal.).
We summed up the number of companion galaxies within 
bins of one magnitude, respectively,
of the H$\alpha$, [OIII], and g-band luminosities. 
We set an upper luminosity limit of $L = 10^{34}\,erg\,s^{-1}$
whenever the emission lines were too weak to be detected in the galaxy spectra
or when the galaxy spectra were too noisy. 
%
%
%
%
%
%

Finally we divided the relative numbers of environmental galaxies
into four separate classes (within the
projected radius of 1 Mpc):\\
- the central galaxy has 0 to 3 companions,\\
- the central galaxy has 4 to 5 companions,\\
- the central galaxy has 6 to 8 companions,\\
- the central galaxy has more than 8 companions.\\

We present in Fig.\,1 the numbers of galaxy companions
as a function of internal H$\alpha$ (Fig.\,1a-c), [\ion{O}{iii}]\,$\lambda 5007$
(Fig.\,1d-f), and
g-band (Fig.\,1g-j) luminosities
for all the different types of central galaxies.
These numbers are summed up within projected
radii of 1 Mpc around the central galaxies.
Given are the relative numbers in percent for the individual 
magnitude bins.


The number of companion galaxies -- within the
projected radius of 1 Mpc of our central galaxies --
is inversely correlated with
the H$\alpha$, [OIII] and continuum luminosities.
This means that the higher the emission line and/or continuum luminosity
of a galaxy, the lower the number of companions.
The same trend appears in all our four galaxy samples (central Sey~1, Sey~2,
HII galaxy, absorption line galaxy) independently.
The given results shown in two Seyfert~1 luminosity bins of Fig.\,1
(log L[OIII]= 37 (Fig\,1.d) and log $L_{cont}$= 41 (Fig\,1.g))
are not significant because they are caused only by one galaxy.
This general trend in Fig.\,1
 is superimposed on the second trend that different
activity types of galaxies lie in different dense environments
(see Table 1).

The number of companion galaxies would decrease by 18 percent when we
defined
physical neighbors by having radial velocities within
$\pm 600\,km\,s^{-1}$  (instead of $\pm 800\,km\,s^{-1}$) 
to the central galaxies. However, the general trends 
remain the same.

\subsection{The activity type of AGN/non-AGN companion galaxies}

We present in Table~3 and
 Fig.~2 the relative numbers of companion galaxies (in percent) belonging to
the different activity types
as a function of 
activity degree of the central AGN/non-AGN galaxy.
The neighbors are divided into
additional subtypes as defined in Section 3.
\begin{table}
\tabcolsep+2mm
\caption{Relative frequency (in percent) of individual activity types in the
neighborhood of
Seyfert, HII, and absorption-line galaxies.}
\centering
\begin{tabular}{r|r|r|r|r|r|r}
\hline 
\noalign{\smallskip}
Central&Sey&LINER&HII&H$\alpha$Em&Abs1&Abs2 \\
gal.   &     &   &   &           &    & \\
\hline 
\noalign{\smallskip}
Sey1 & 19.7 & 1.2 & 38.3 & 14.2 & 6.0 & 20.6 \\
Sey2 & 15.1 & 0.8 & 46.0 & 11.6 & 6.3 & 20.2 \\
HII & 14.0 & 0.5 & 50.8 & 10.9 & 5.6 & 18.3 \\
Abs.&  8.9 & 0.5 & 24.4 & 10.2 & 8.9 & 47.2 \\
\noalign{\smallskip}
\hline 
\end{tabular}
\end{table}

HII galaxies are the most common type except in the environment of
absorption-line galaxies. Pure LINER spectra are rarest.
There is a clear trend by a neighborhood effect within the projected radius of
1 Mpc toward the AGN and non-AGN types: Seyfert galaxies tend to have
the highest probability of having another Seyfert galaxy in the
 neighborhood. HII galaxies tend to have
the highest probability of having another HII galaxy in the
 neighborhood. And finally, absorption-line galaxies tend to have
the highest probability of having another absorption-line galaxy in their
 neighborhood.

\subsection{H$\alpha$ and [\ion{O}{iii}]\,$\lambda 5007$ emission
line luminosities
in the spectra of AGN/non-AGN companions}

The H$\alpha$ and [OIII] line luminosities in the environment
of Seyfert~1, Seyfert~2, HII, and absorption line galaxies
are shown
in Figs. 3 and 4 as a function of projected distance to the central galaxy.
 We are
studying whether emission line intensities in neighboring galaxies
are a function of distance to the central object and whether this relation
depends on the activity degree of the central galaxy.   
The line intensities in these figures
 are given on a logarithmic scale. 
%
%
%
%
\begin{figure*}
\hbox{
\includegraphics[width=56mm,height=85mm,angle=270]{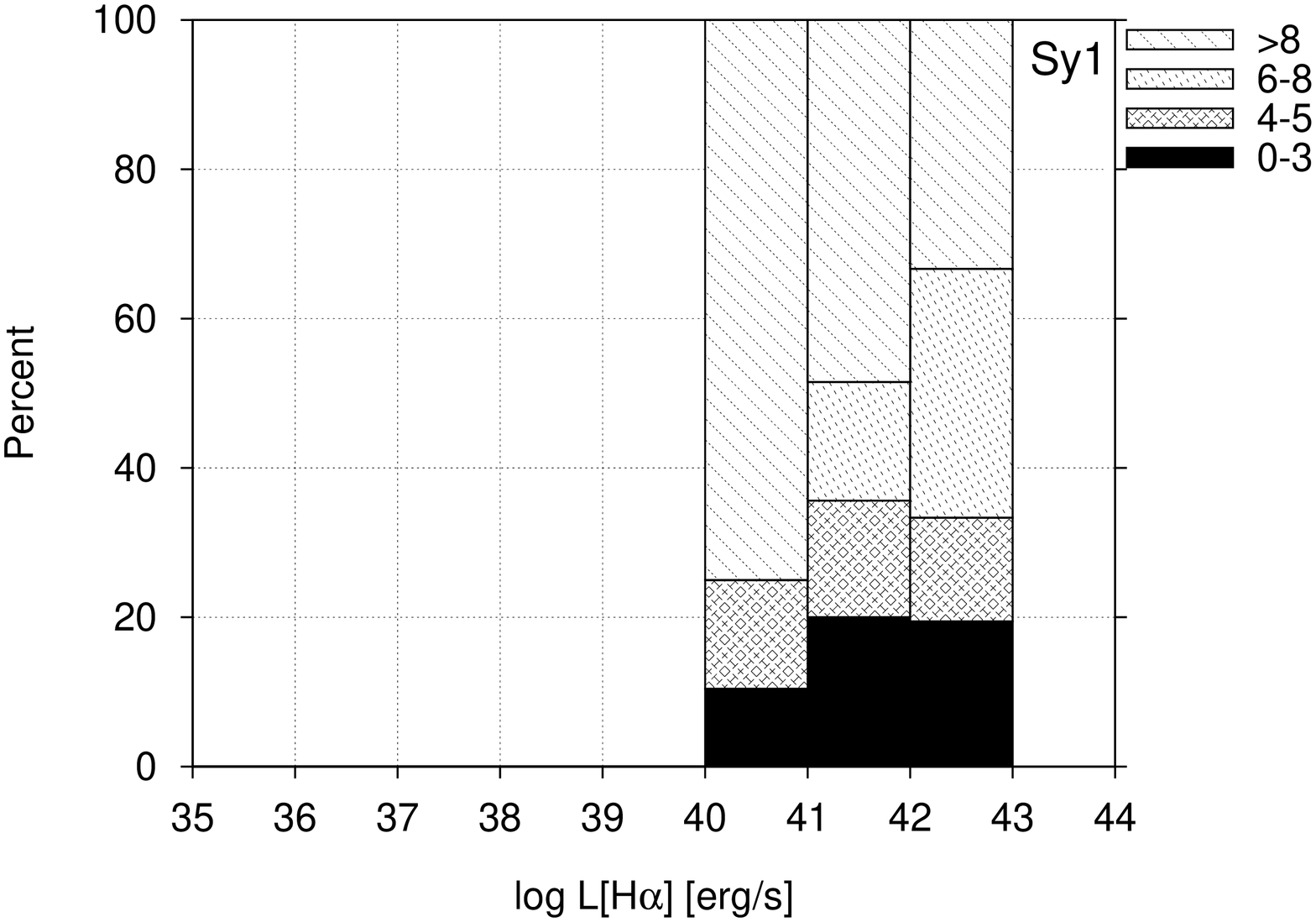}
\hspace*{7mm}      
\includegraphics[width=56mm,height=85mm,angle=270]{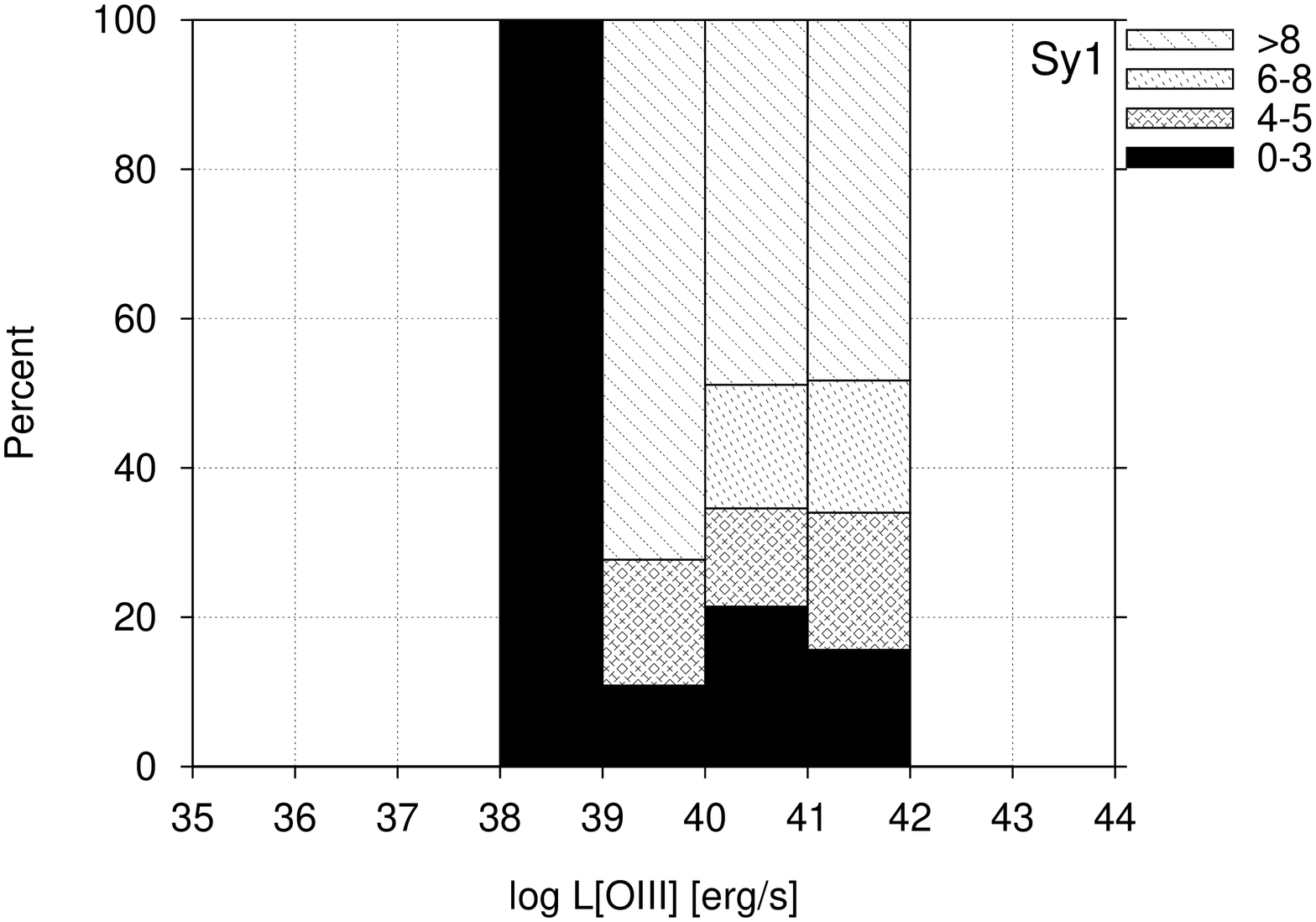}
\hspace*{7mm}
}
\hbox{
\includegraphics[width=56mm,height=85mm,angle=270]{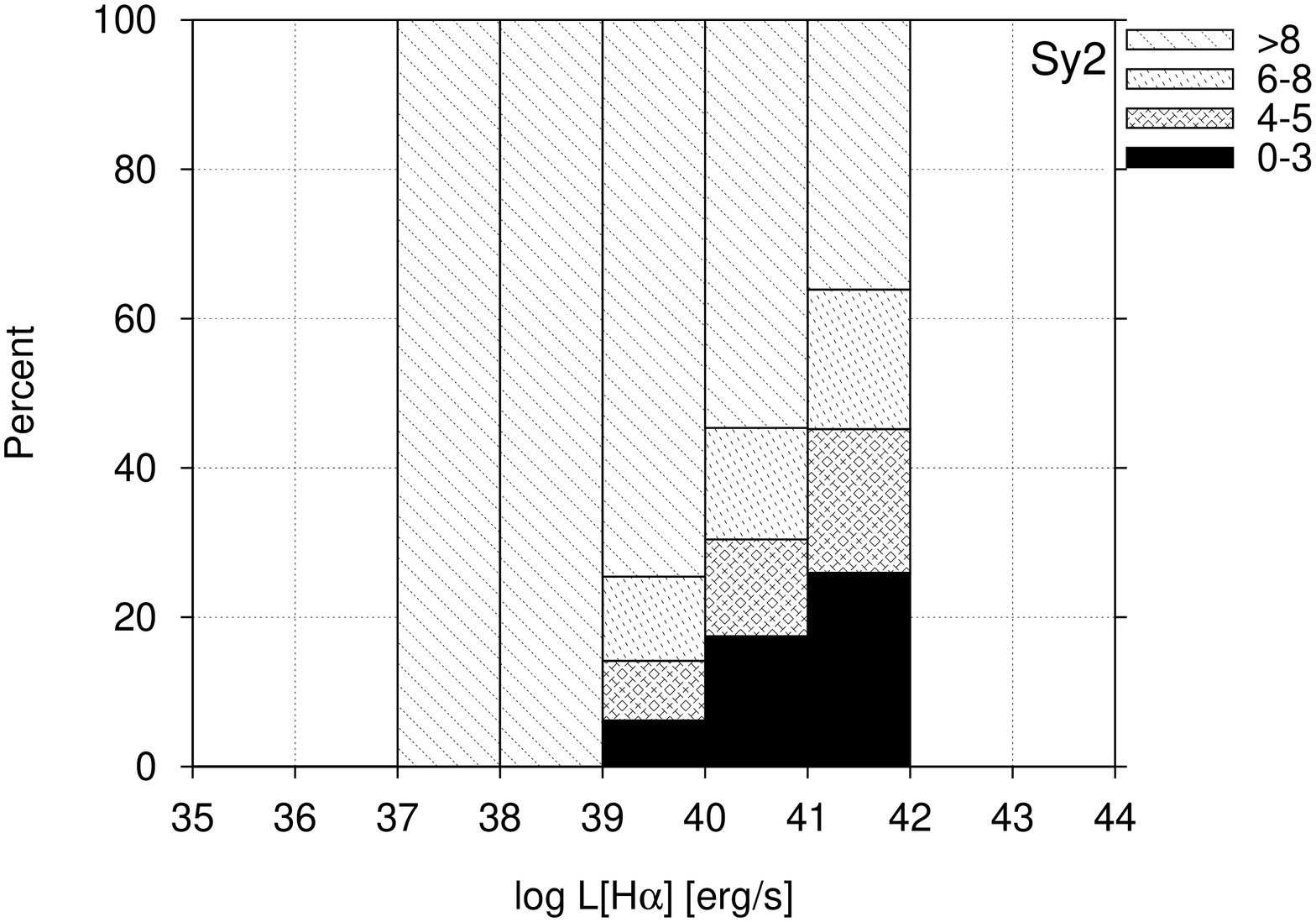}\hspace*{7mm}  
\includegraphics[width=56mm,height=85mm,angle=270]{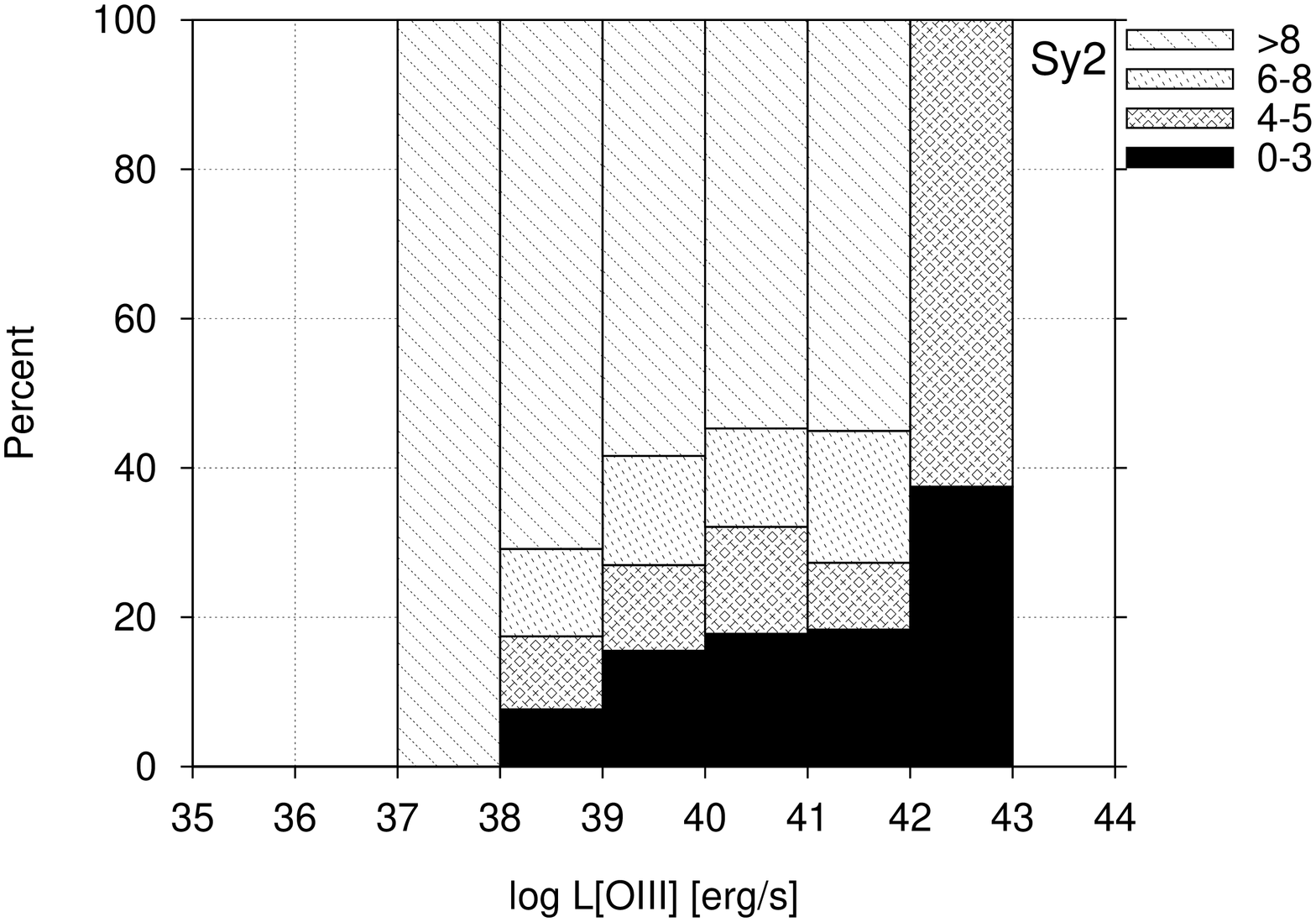}\hspace*{7mm}}
\hbox{
\includegraphics[width=56mm,height=85mm,angle=270]{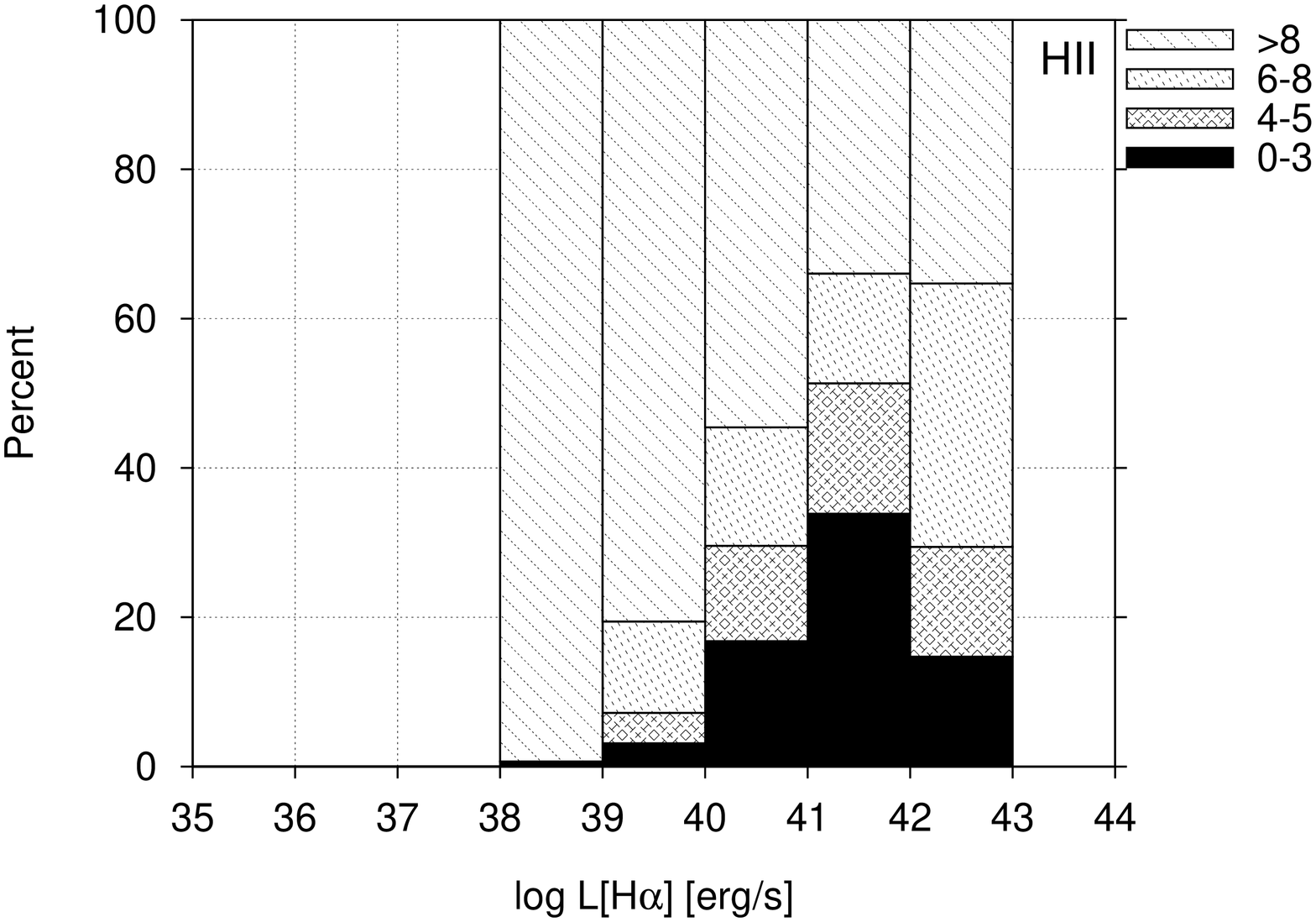}\hspace*{7mm}    
\includegraphics[width=56mm,height=85mm,angle=270]{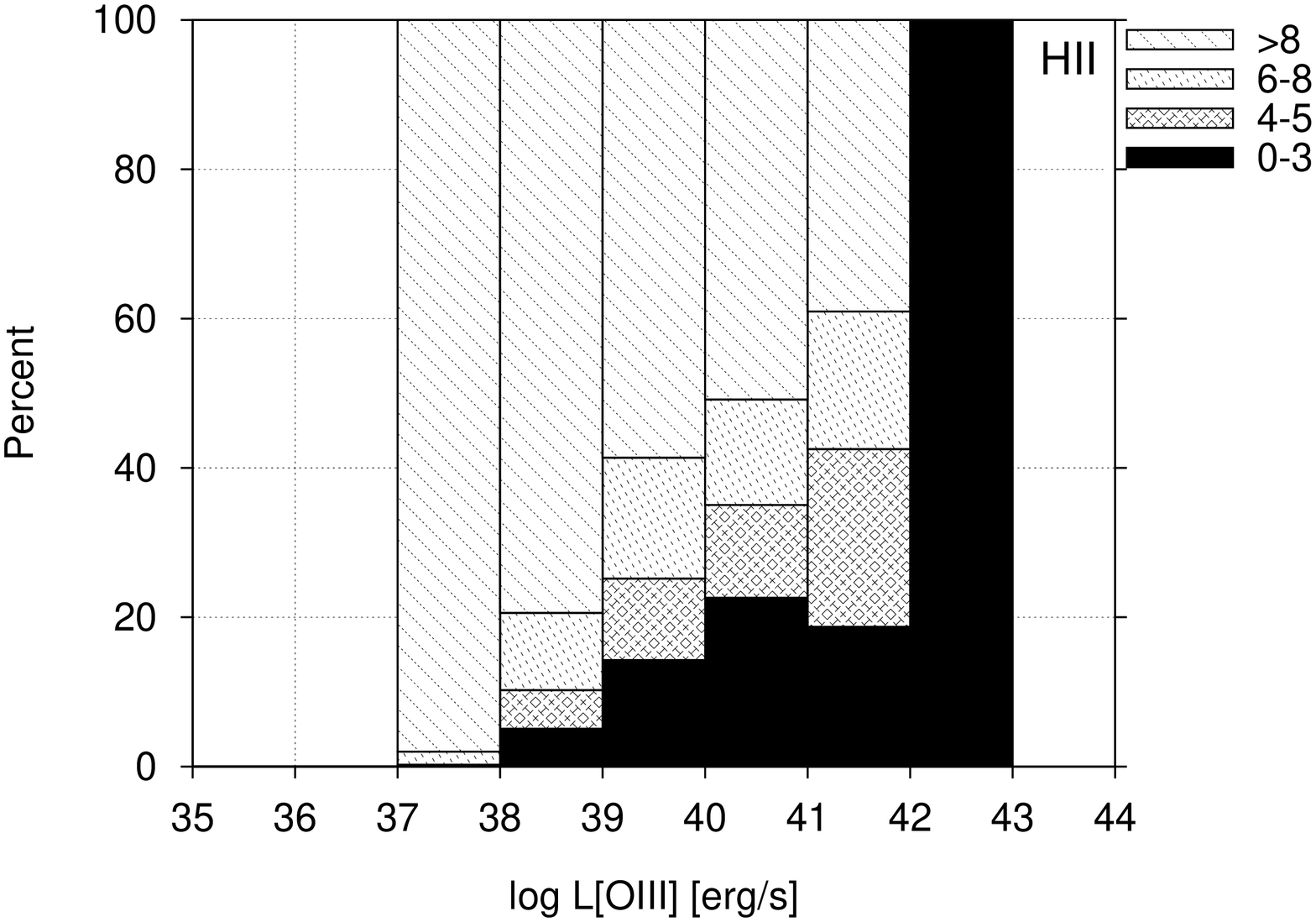}\hspace*{7mm}}
  \caption*{Fig. 1 a-f Relative number of neighboring galaxies 
(divided into four companion classes: 0-3, 4-5, 6-8, $>$8 companions)
as a function of the
H$\alpha$ and [\ion{O}{iii}]\,$\lambda 5007$ line luminosities  
(divided into luminosity bins) of the central Seyfert~1, 2,
and HII galaxies.}
\end{figure*}

The mean H$\alpha$ and [\ion{O}{iii}]\,$\lambda 5007$ luminosities of
the central AGN/non-AGN galaxies, along with 
 of their environmental
galaxies within 1 Mpc, are presented in Table 4. Those
galaxies showing only upper H$\alpha$ and [OIII] line
luminosity limits of $L = 10^{34}\,erg\,s^{-1}$  in their spectra
were not considered
for the determination of the mean values.
The central Seyfert~1 galaxies exhibit the highest
 mean H$\alpha$ and [OIII] line luminosities
in their spectra. Seyfert~2 and HII galaxies
emit  H$\alpha$ and [OIII] lines that are one magnitude
fainter on average.

Considering the environmental galaxies there are two general trends.
The environmental galaxies around both Seyfert~1 and Seyfert~2 
galaxies emit the highest
internal  H$\alpha$ and [\ion{O}{iii}]\,$\lambda 5007$ line luminosities.
Seyfert~1 environmental galaxies exhibit mean luminosities of
$log\,L\,=\,39.64\,erg\,s^{-1}$ in H$\alpha$ and
$log\,L\,=\,38.94\,erg\,s^{-1}$ in the O[III] line.
The mean line luminosities of the companions are lower 
for Seyfert~2 and HII galaxies.
%
\begin{figure}[!h]
\hbox{
\includegraphics[width=56mm,angle=270]{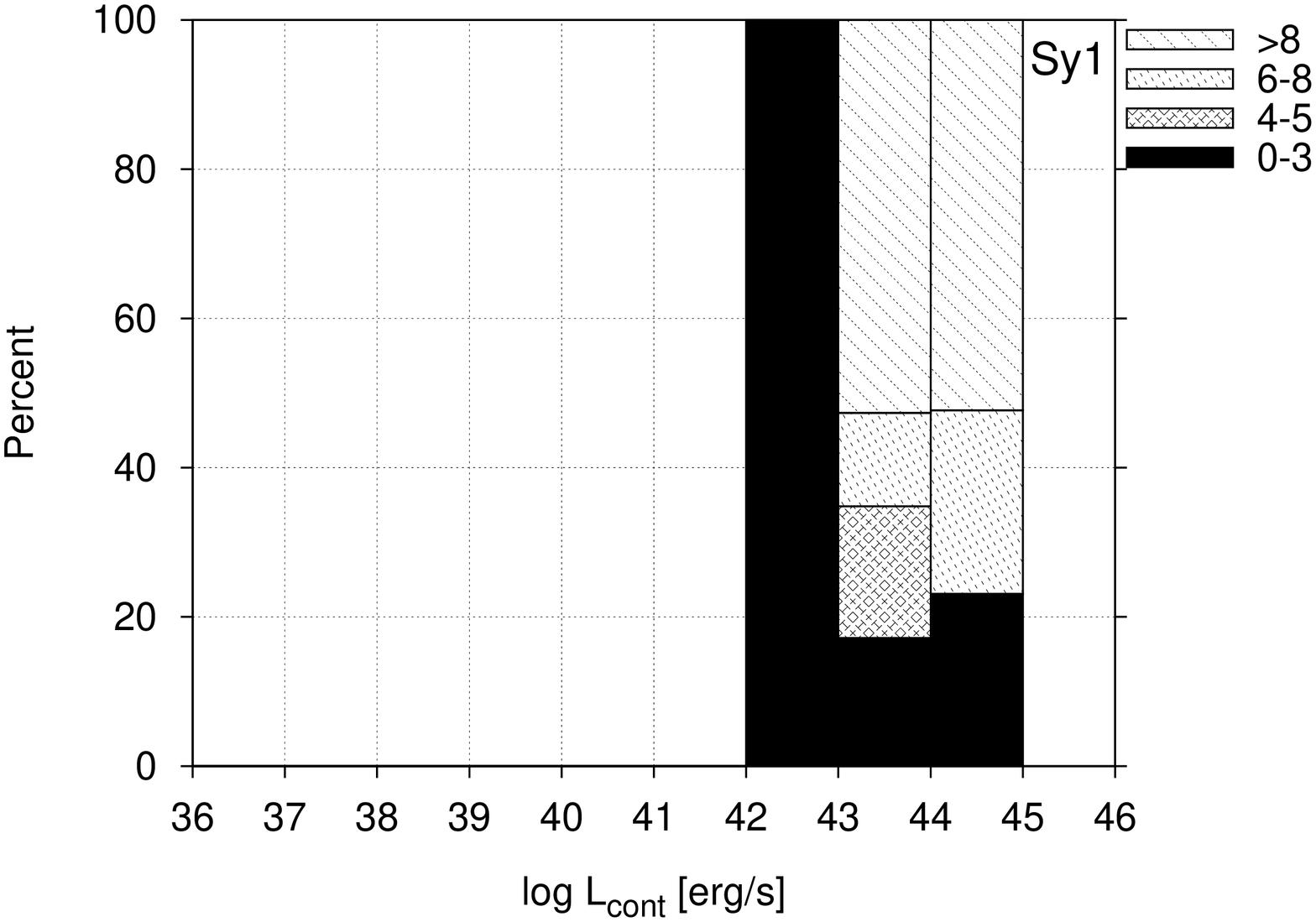}\hspace*{-7mm}      
}
\hbox{
\includegraphics[width=56mm,angle=270]{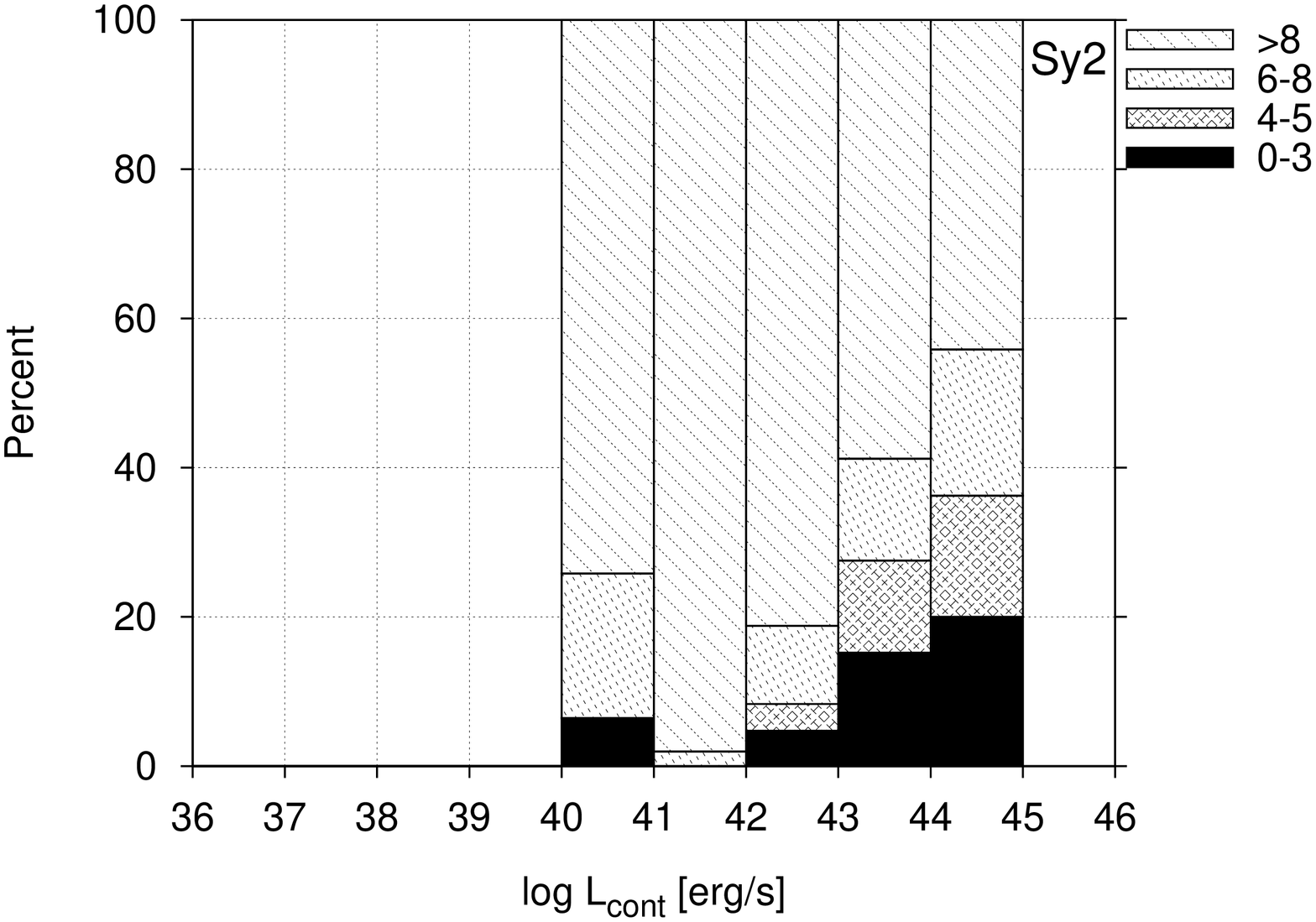}\hspace*{-7mm}    
}
\hbox{
\includegraphics[width=56mm,angle=270]{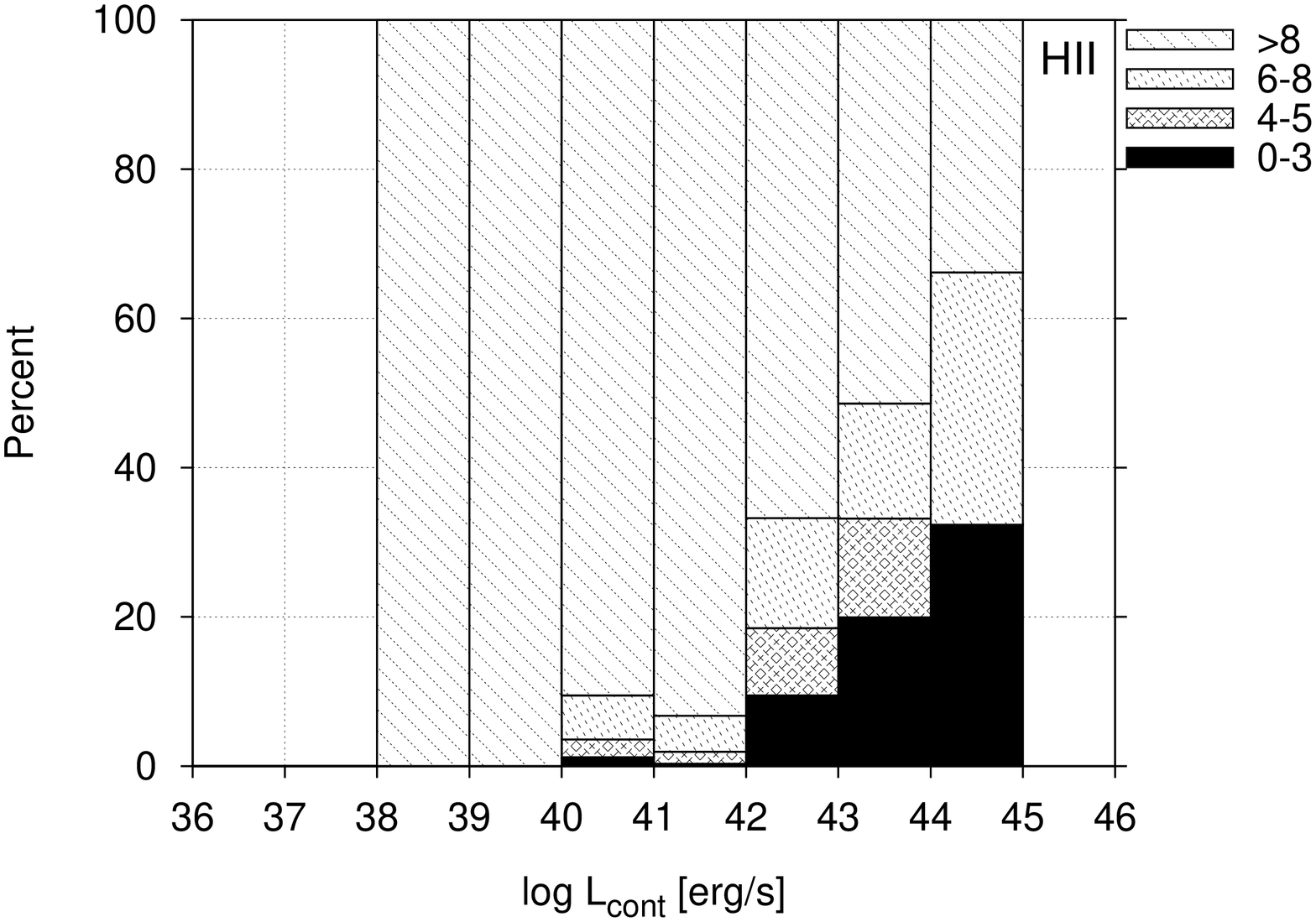}\hspace*{-7mm}  
}
\hbox{
\includegraphics[width=56mm,angle=270]{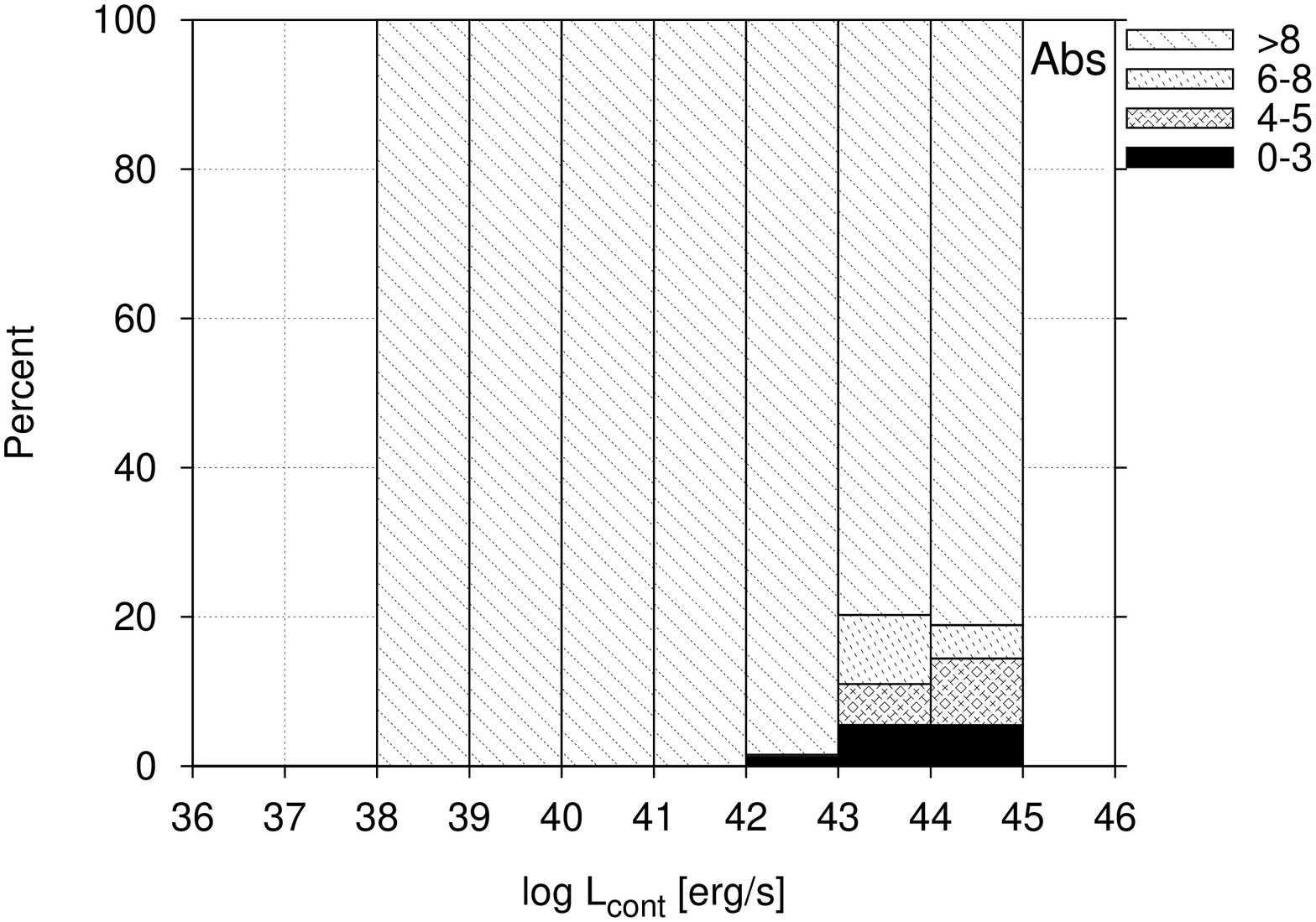}\hspace*{-7mm}      
}
\caption*{Fig. 1 g-j.  Same as Fig. 1 a-f: Relative number of
neighboring galaxies
for g-band continuum luminosity bins of the  central Seyfert~1, 2,
HII, and absorption line galaxies.}
\end{figure}
%
%
%
\begin{figure}
\includegraphics[width=8.0cm,angle=0]{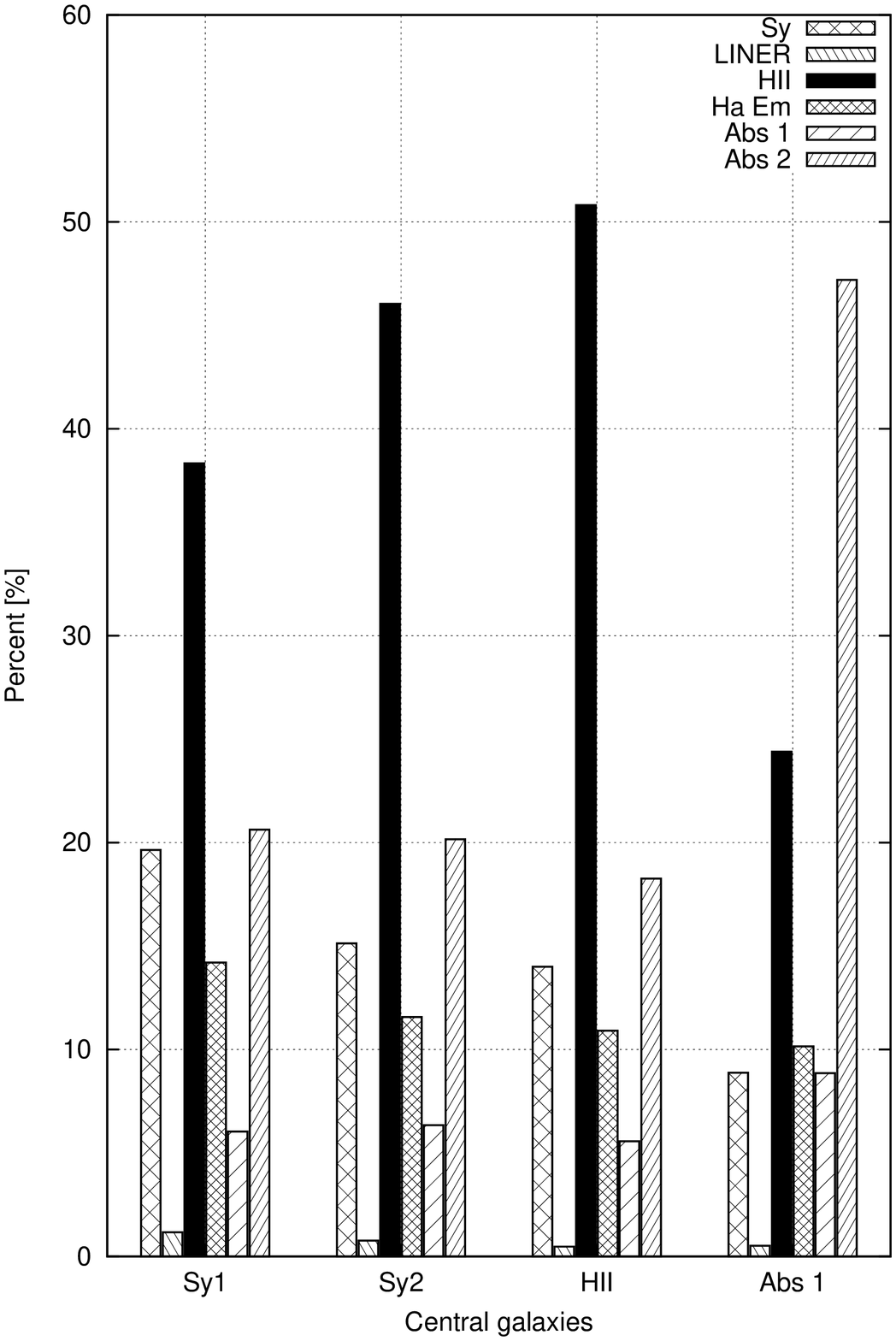} 
       \vspace*{5mm} 
  \caption{The spectral activity type of neighboring galaxies as a
function of the 
activity degree of the central galaxy.}
   \label{actdegrneigh.eps}
\end{figure}
%
%
The lowest mean H$\alpha$ and [OIII] line luminosities of
$log\,L\,=\,39.13\,erg\,s^{-1}$ in H$\alpha$  and
$log\,L\,=\,38.14\,erg\,s^{-1}$ in the O[III] line exhibit the companions
of the central absorption line galaxies.

The second result is to be seen in the line luminosities
of the environmental galaxies considering the
projected distance to the central AGN/HII galaxy.
There is a trend toward the H$\alpha$ and [OIII] luminosities
in the environmental galaxies
becoming stronger when they are located closer to
the central emission line galaxy. 
This observation is even
diluted by the fact that we consider neighborhood galaxies at projected
distances from the central galaxies.
Galaxies with small projected distances to the central galaxy
might be at large background or foreground distances
in reality.
Environmental galaxies around absorption line galaxies
show the opposite trend
to the central Seyfert and HII galaxies: Here the H$\alpha$
and [OIII] luminosities of the
environmental galaxies are increasing as a function of projected distance
to the central absorption line galaxy. This is evidence for the
morphology density relation that has been mentioned in the introduction.

Figure 5 again demonstrates the trend toward
 increasing/decreasing activity by the
companion galaxies as a function of projected distance to the central galaxy.
Here the line intensities are plotted
on a linear intensity scale. The line intensities are
binned over intervals of 50 kpc projected distance. The companions of the
Seyfert~1 galaxies are not considered in these figures since their quantity
is statistically not that significant. One can see the clear trend for the mean
H$\alpha$ and [OIII] luminosities of the Seyfert~2 and HII companions 
to increase toward the central galaxy by a factor of two to four.
Again one has to consider the dilution factor of the central foreground and
background galaxies.
The increasing/decreasing activity begins at projected
 distances of 200 -- 400 kpc. The rise is
steeper for HII galaxies than for Seyfert~2 galaxies.
The central absorption line galaxies show the opposite trend.
Here the mean line
intensities of the companions increase by a factor of two over a
distance of 1 Mpc. 

The log of the mean
H$\alpha$ and [OIII] luminosities in the environmental
galaxies is listed in Table 5. It is given for different shells
at distances of 0.-0.1 Mpc, 0.1-0.2 Mpc, 0.4-0.6 Mpc, as well as 0.8-1.0 Mpc.

%
\begin{figure*}
 \hbox{
\includegraphics[width=11.2cm,angle=270]{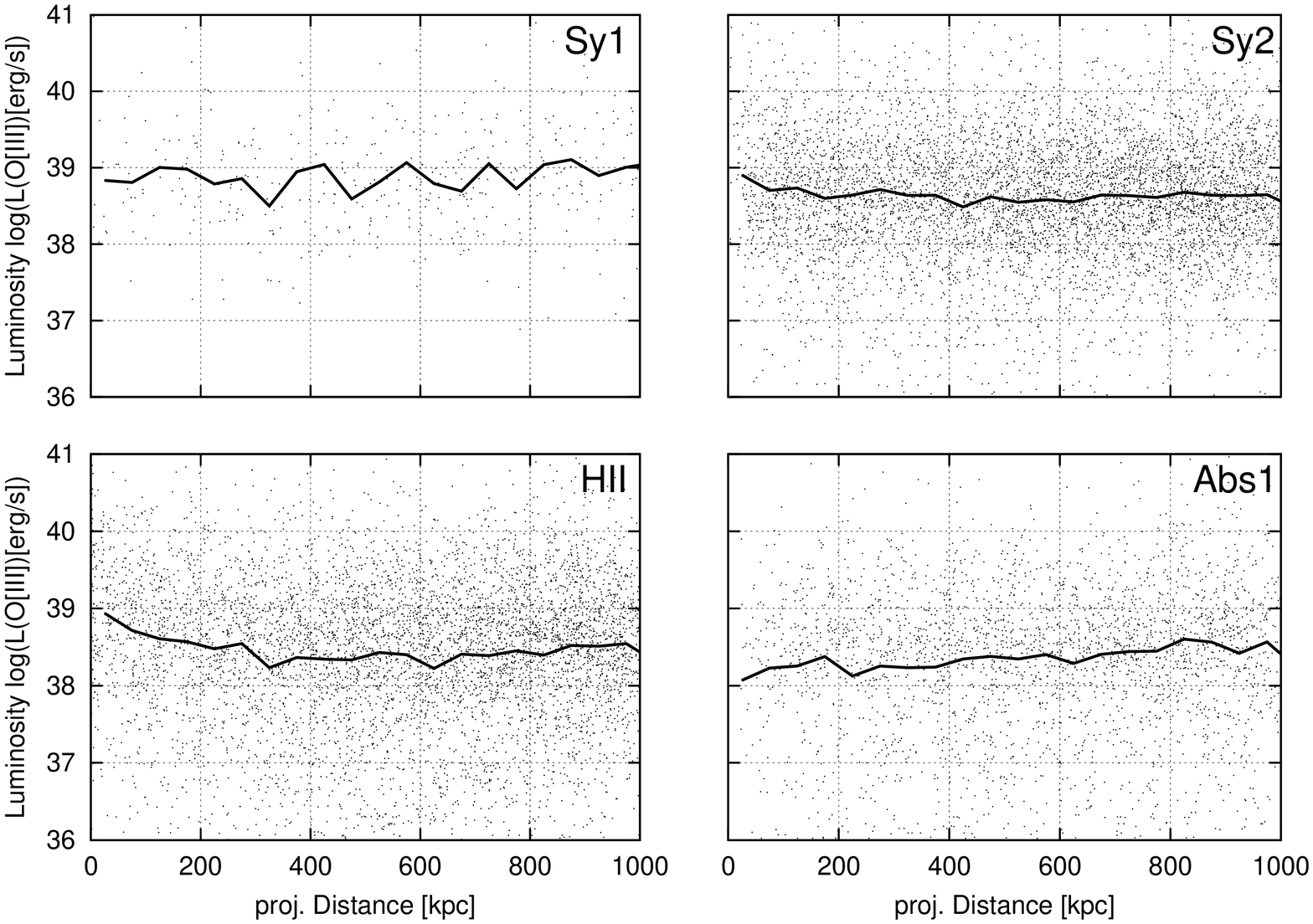}
}
  \caption{[\ion{O}{iii}]\,$\lambda 5007$ luminosity of the companion galaxies as a function of
the distance to the central galaxy. The thick solid line shows
the mean line intensities
binned over intervals of 50 kpc projected distance.} 
   \label{figlum03.eps}
\end{figure*}
%
\begin{figure*}
 \hbox{
\includegraphics[width=11.2cm,angle=270]{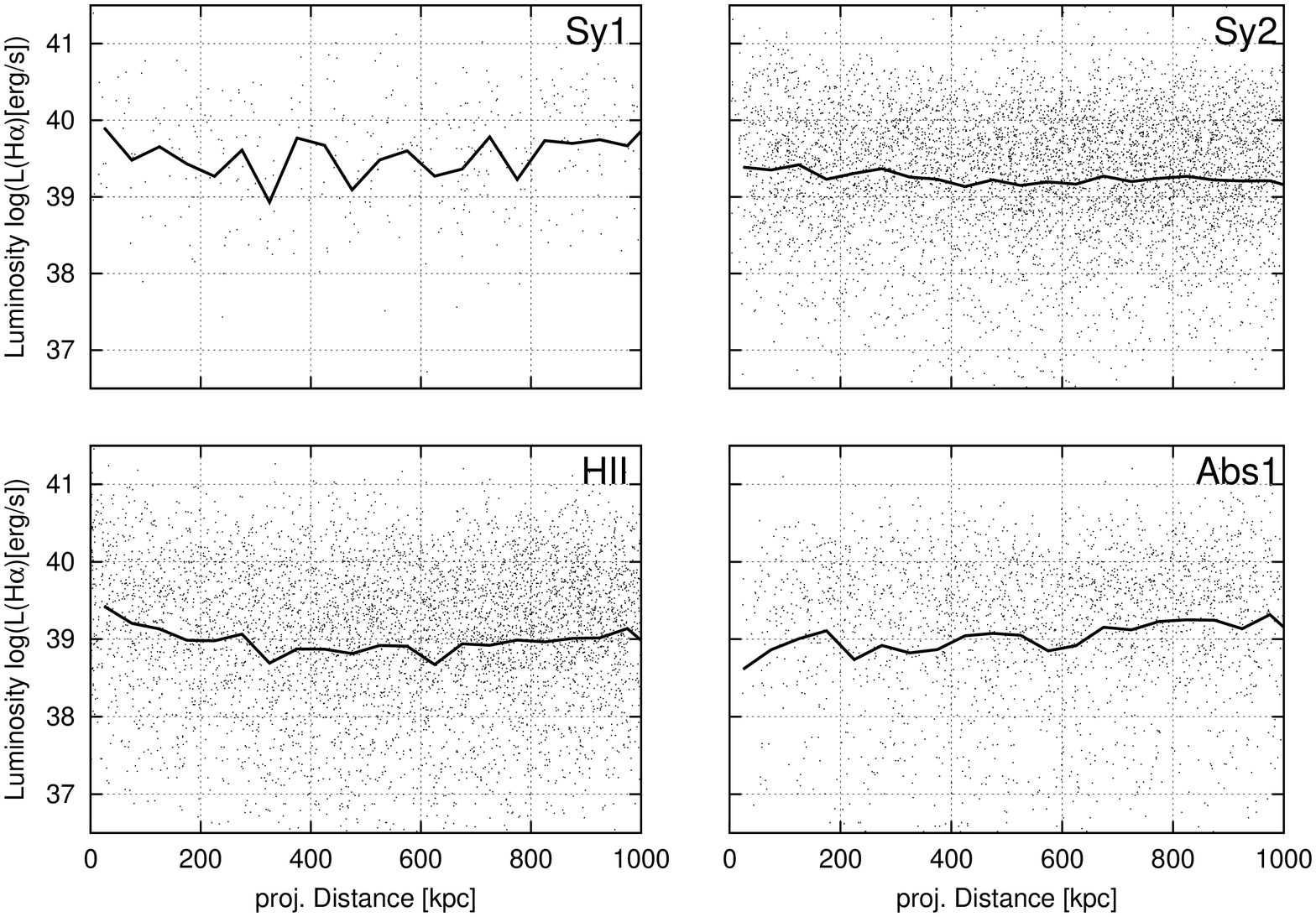}
}
  \caption{H$\alpha$ luminosity of the companion galaxies as a function of
the distance to the central galaxy.}
   \label{figlumha.eps}
\end{figure*}
%

%
%
%
\begin{figure*}
\hbox{\includegraphics[clip,width=55mm,height=85mm,angle=270]{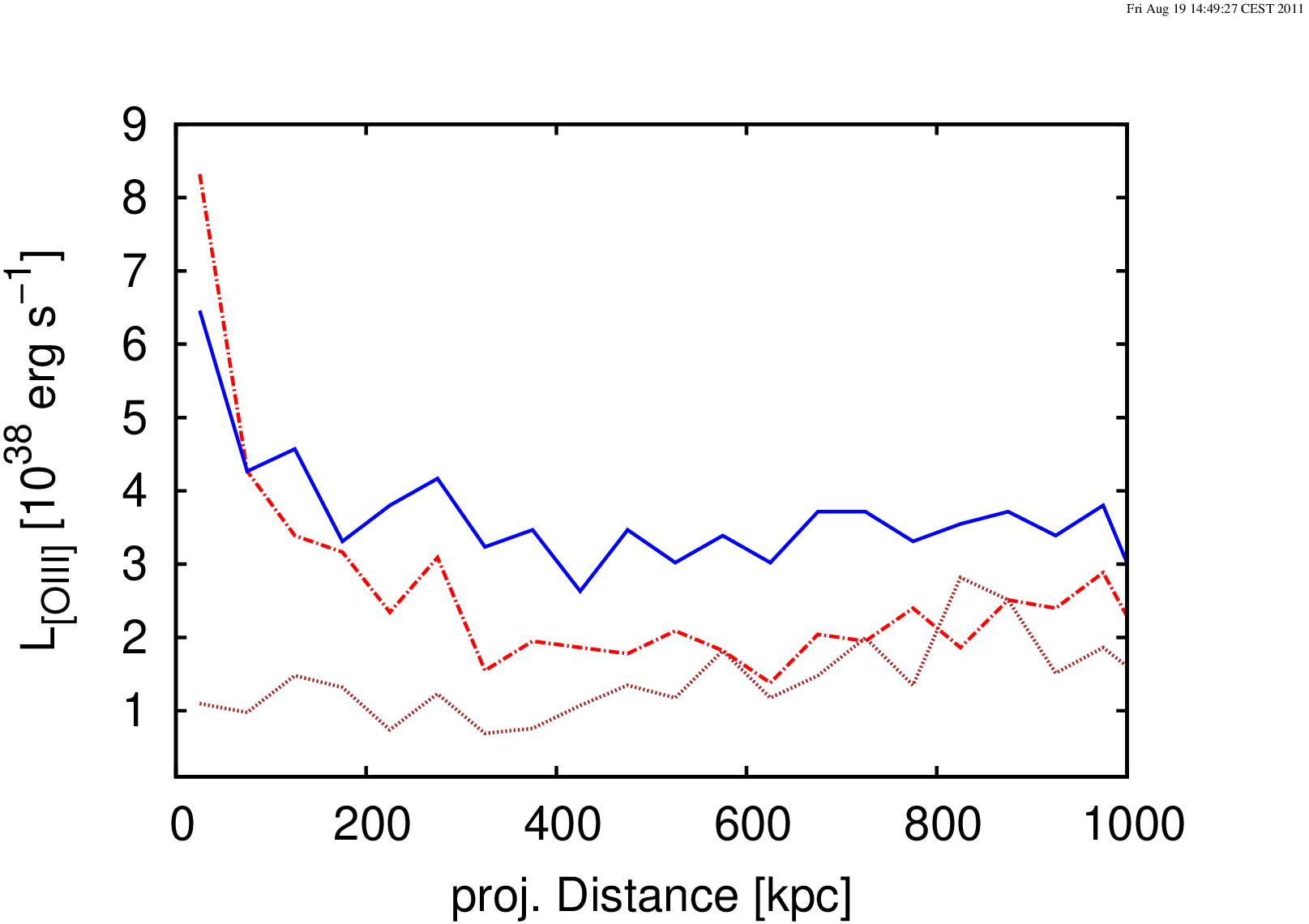}\hspace*{0mm}      
\includegraphics[clip,width=55mm,height=85mm,angle=270]{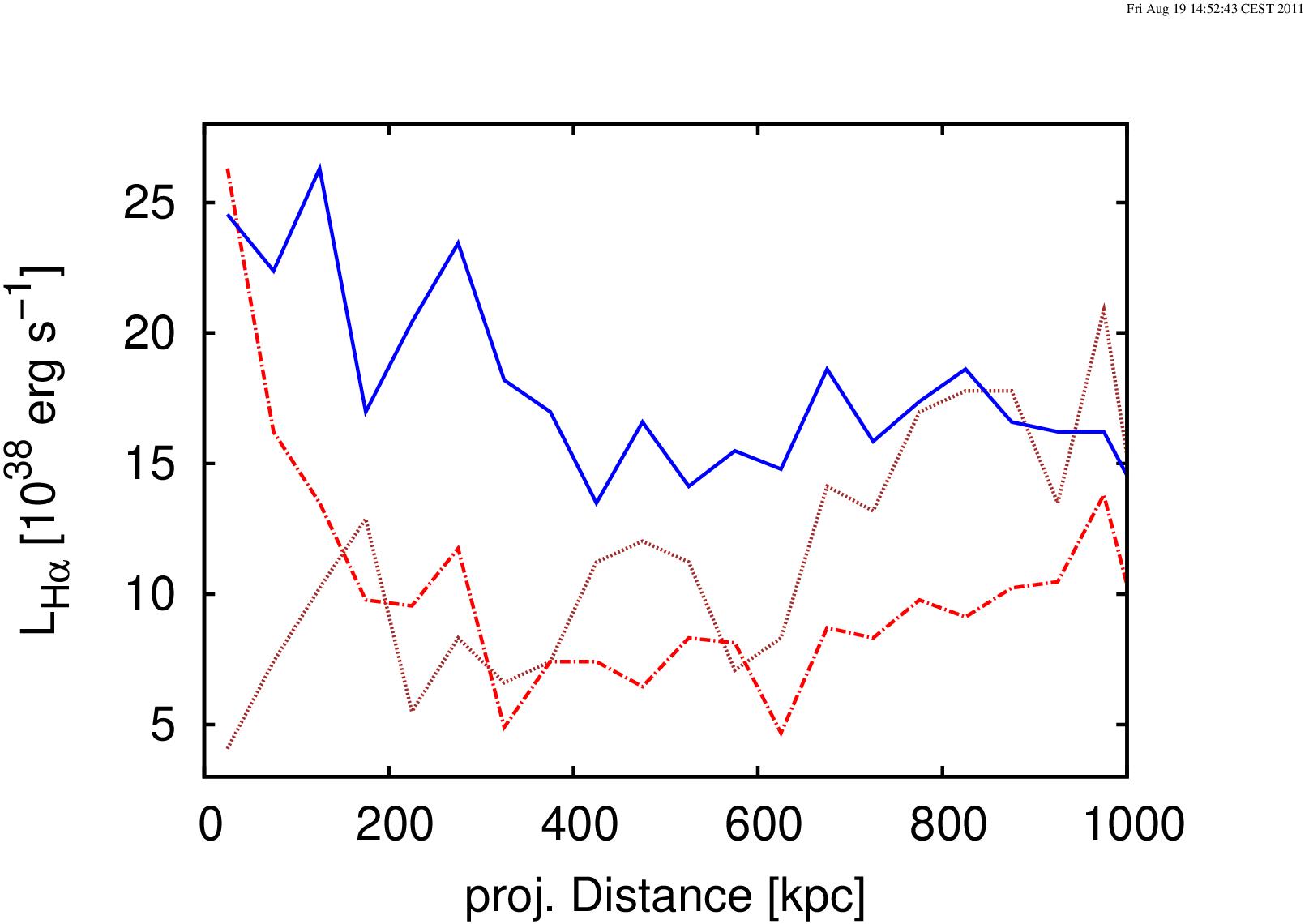}}
  \caption{Comparison of the [OIII] and H$\alpha$ luminosities
 in the environmental galaxies
 around central Seyfert~2 (blue solid line), HII (red dot-dashed line), and
 absorption galaxies (brown dotted line) as a function of distance to the
central galaxy.}
\end{figure*}
%
%

The trend toward increasing/decreasing line luminosity as a function of
distance does not apply to the continuum luminosities of the companion
galaxies. Figure\,6 shows the g-band continuum luminosities of the Seyfert~2 and
 HII companion galaxies as a function of
the distance to the central galaxy. The definition of the magnitude $\mu$
is given by
\[ \mu(x) = -a [ sinh^{-1}(x/2b) + ln\enspace b ]\]
where x is the dimensionless normalized flux $x = f/f_{0}$ and
$a= 2.5/ln(10)= 1.08574$. The ``softening'' constant b 
determines the flux level at which linear behavior sets in (see 
Lupton et al. \cite{lupton99} for details).
       
 The relative g-band continuum luminosities of the Seyfert~2 and HII companion
galaxies
are independent of the distance to the central galaxy
in contrast to the H$\alpha$ and [OIII] line luminosities.
 The other Sloan continua bands show
the same constant characteristics as the g-band continuum.

\section{Discussion and conclusion}

We investigated the spectral properties of the surrounding
galaxies around central Seyfert/non-Seyfert galaxies
in a large-scale projected environment of 1 Mpc. We restricted ourselves
to galaxies having redshifts z of less than 0.08.
Altogether, we analyzed
the spectra of about 25,\,000 spectra from the Sloan Digital Sky Survey (SDSS). 


\begin{table}
\tabcolsep+2mm
\caption{Log of mean H$\alpha$ and [\ion{O}{iii}]\,$\lambda 5007$ line
 luminosities (erg\,s$^{-1}$) 
 of the central galaxies,
as well as of the environmental galaxies
within 1000 kpc.}
\centering
\begin{tabular}{c|c|c|c|c}
\hline 
\noalign{\smallskip}
activity type & L(H$\alpha$) & L(H$\alpha$) &  L(O[III]) &  L(O[III]) \\
of centr. gal.& central      & compan.      & central    &  compan.  \\   
\hline 
\noalign{\smallskip}
 Sey1 &  41.48 &  39.64 & 40.64 & 38.94 \\
 Sey2 &  40.45 &  39.30 & 39.80 & 38.63 \\
 HII &  40.61 &  39.09 & 39.89 & 38.52 \\
 Abs.&    -   &  39.13 &  -    & 38.14  \\
\noalign{\smallskip}
\hline 
\end{tabular}
\end{table}
%

%
%

%
%
\begin{figure*}
 \hbox{
\includegraphics[width=11.3cm,angle=270]{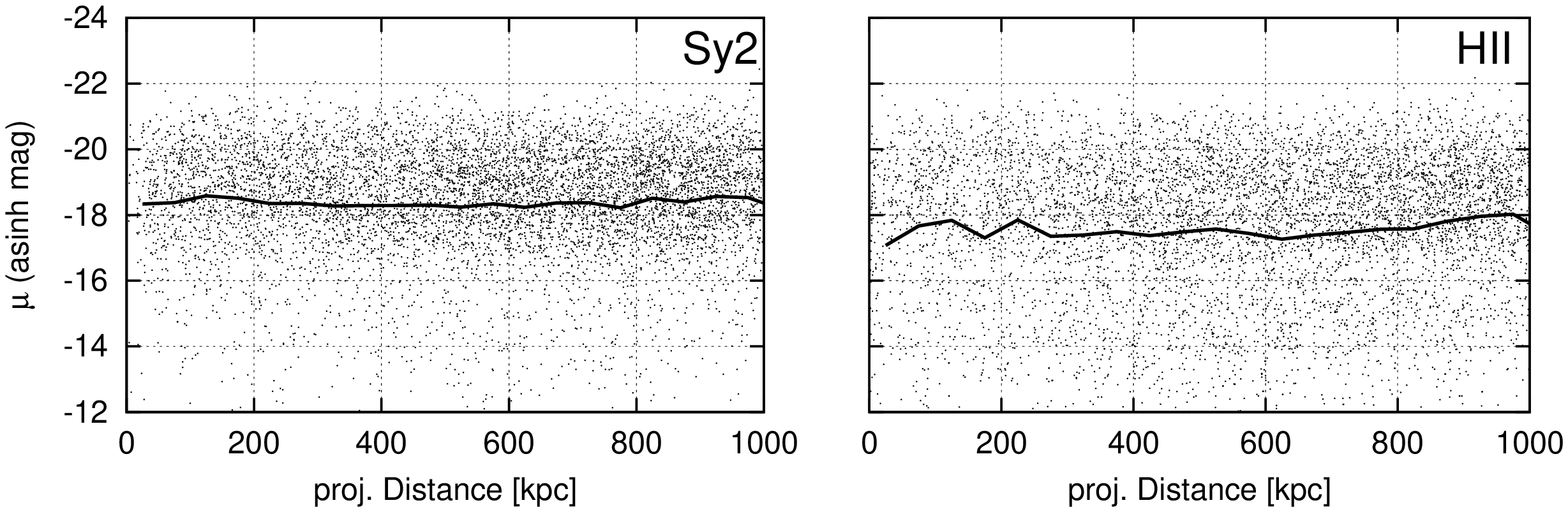}
}
       \vspace*{-58mm} 
  \caption{g-band continuum luminosity of the Seyfert~2 and HII companion galaxies
 as a function of distance to the central galaxy.} 
   \label{helligkeitabstand-h2-sy2-proj.eps}
\end{figure*}
%
%

%
\begin{table}
\tabcolsep+1mm
\caption{Log of mean [\ion{O}{iii}]\,$\lambda 5007$ and  H$\alpha$ line luminosities (erg~s$^{-1}$) 
 in the environmental galaxies given for different types of central galaxies
 and for different spherical shells at different distances.
}
\centering
\begin{tabular}{c|c|c|c|c}
\hline 
\noalign{\smallskip}
act.type& L(O[III]) & L(O[III]) &  L(O[III]) &  L(O[III]) \\
centr.gal.& 0--0.1 Mpc & 0.1--0.2 Mpc & 0.4 --0.6 Mpc & 0.8--1.0 Mpc\\   
\hline 
\noalign{\smallskip}
 Sey1 &  38.89 &  38.79 & 38.76 & 38.98 \\
 Sey2 &  38.70 &  38.55 & 38.49 & 38.51 \\
 HII &  38.69 &  38.44 & 38.25 & 38.35 \\
 Abs.&  38.07 &  38.00 & 38.11 & 38.28  \\
\noalign{\smallskip}
\hline 
\noalign{\smallskip}
act.type& L(H$\alpha$) & L(H$\alpha$) &  L(H$\alpha$) &   L(H$\alpha$) \\
centr.gal.& 0--0.1 Mpc & 0.1--0.2 Mpc  & 0.4 --0.6 Mpc  &  0.8--1.0 Mpc \\  
\hline 
\noalign{\smallskip}
 Sey1 &  39.68 &  39.35 & 39.42 & 39.78 \\
 Sey2 &  39.38 &  39.27 & 39.17 & 39.20 \\
 HII &  39.25 &  38.98 & 38.84 & 38.99 \\
 Abs.&  38.83 &  38.92 & 38.99 & 39.19  \\
\noalign{\smallskip}
\hline 
\end{tabular}
\end{table}

\subsection{The number of companions as function of internal
emission line and continuum intensities}

We analyzed the environmental galaxies around 1,\,594
central AGN (47 percent), around 1,\,406 central HII galaxies (41 percent),
and around 415 central absorption line galaxies (12 percent).
The relative numbers of these central activity types are similar to those of a
complete sample of nearby galaxies consisting of 486 galaxies
(Ho et at., \cite{ho97}). They identified
15 percent of their nearby galaxies as absorption line galaxies,
40 percent as HII nuclei, and
45 percent as AGN (including LINERs) (their Table 1).

We demonstrated that the number of companion galaxies
is inversely correlated to the internal
H$\alpha$ and [\ion{O}{iii}]\,$\lambda 5007$ line,
as well as to continuum luminosities
in projected environments of 1 Mpc around a central galaxy (Fig.~1).
This universal trend is seen
in all our samples independent of the activity degree of the central galaxy.
The general results remained the same when we chose subsamples 
consisting of e.g. central galaxies with g-band magnitude limits of only 17.

Schmitt (\cite{schmitt01}) has investigated the
immediate galaxy density distribution
among galaxies of different activity types
in an earlier study. He considered secondaries
at distances five times the diameter of the primary.
He found more immediate
neighbors near absorption line galaxies.
In the larger environment of absorption line galaxies,
we detected more companions as well.
Schmitt (\cite{schmitt01})
explains his finding with the morphology-density effect of different
morphological galaxy types. However, Constantin \& Vogeley (\cite{constantin06})
carried out
 an independent investigation of the clustering of low-luminosity AGN.
They also find that Seyfert galaxies are less clustered than normal galaxies.
They claim that the clustering around AGN
 is not driven in a simple way by the morphology-density relation,
since colors and concentration indices follow similar distributions.

Miller (\cite{miller04}) investigated the effect of galaxy environment on
observed H$\alpha$ emission line intensities in galaxies to derive their
star-formation rate and AGN activity based on an early release of the SDSS.
In analogy to the density-morphology relation, he presented a
density-morphology-star formation relation. He found that the fraction
of star-forming galaxies decreases in dense regions while the fraction of
AGN is constant over all densities. He explains this result by the
idea that the AGN phenomenon is mainly related to the bulge
component of galaxies and does not depend to any great extent  
on the galaxy environment. 

The cause for a general relation between emission line intensities
and a lower number of companion galaxies
should be a topic for future  investigations.
 One idea for explaining this relation
might be the limited
amount of intergalactic gas that streams
into the galaxies and leads to the formation of emission lines. This idea is
similar to the picture of Dekel et al. (\cite{dekel09}) that cold streams 
in massive halos are the main mode of galaxy formation.

The star formation rate in galaxies is driven partly by internal processes (gas
consumption) and partly by local or large-scale environmental effects.
Diaferio et al. (\cite{diaferio01})
 discusses the concept that galaxies in groups or clusters
that are not located at the center of their potential are assumed
to have no hot gas reservoir for refueling the gas for star formation activity. 

 Wake et al. (\cite{wake04}) present two-point correlation functions (2PCF)
of narrow-line AGNs. They show that their two-point correlation function
depends on the [\ion{O}{iii}] line luminosities in their AGN. 
Low-luminosity AGNs have a higher clustering amplitude than high-luminosity
 AGNs. They consider their finding with the idea that low-luminosity AGNs 
reside in more massive galaxies and that the luminosity is an indicator for the 
fueling rate. In hierarchical models of structure formation, more massive dark
matter halos are more strongly clustered.

The lower number of companion galaxies
is inversely correlated to the continuum luminosities of the central galaxies
as well.
The continuum luminosity correlates very roughly
to the mass of the central galaxy.
This finding means that, on average, more massive galaxies are surrounded
by fewer companion galaxies.

\subsection{The number of AGN companions}

Numerous studies about the number of AGN/QSO companions have been executed
in the past. There are indications in some investigations that the
AGN environment depends, among others, on the investigated scale around
the AGN, on the
cosmological distance of the objects, as well as on the AGN type.
 Miller et al. (\cite{miller03}) and Coldwell \& Lambas 
(\cite{coldwell06}) analyzed the environment of a few thousand nearby AGN 
(z$<$0.2). They find similar number densities around AGN and local galaxies.
Li et al. (\cite{li06}) analyzed the clustering of narrow-line AGN
(Kauffmann et al., \cite{kauffmann03}) in the local Universe too.
 They claim that
on scales larger than a few Mpc, their narrow-line AGN have almost the same
clustering amplitude as their control sample of inactive galaxies. On scales
between 100 kpc and 1 Mpc their AGN are more weakly clustered, while they are
marginally more strongly clustered on scales less than 70 kpc.
Shirisaki et al. (\cite{shirisaki11}) studied the AGN-galaxy
clustering of 1809 AGN at redshifts from 0.3 to 3.0 using
Subaru Suprime-Cam images and UKIDSS catalog data.
At higher redshifts they find a
significant excess of galaxies around AGN, while for lower redshifts (z$<$0.9) 
AGN resided in similar environments to the
typical local galaxies. On the other hand,
they detected that the majority of galaxies that are observed to be clustered
around the AGN are blue star-forming galaxies.
Based on 1800 nearby AGN
Sorrentino et al. (\cite{sorrentino06}) see no difference in
the large-scale environment of Seyfert~1 and Seyfert~2 galaxies.

However, we found in our study (see Table 1)
that Seyfert~1 galaxies have on average
fewer companion galaxies than Seyfert~2 or HII galaxies within an
environment of 1 Mpc.
In a similar spirit, Koulouridis et al. (\cite{koulouridis06}) uncovered
the trend for
the fraction of Seyfert~2 galaxies with a close
neighbor to be significantly higher than for Seyfert~1 galaxy samples
within projected distances of 100 kpc. They
explained the difference by different morphologies of their host galaxies. 
Strand et al. (\cite{strand08}) explored the environment of AGN with respect
to redshift, type, and luminosity. They find an increased overdensity of
Seyfert~2 galaxies compared to Seyfert~1 galaxies on scales out to 2 Mpc in
overlapping redshift ranges.

Seyfert~2 galaxies have more companions than Seyfert~1 galaxies
does not accord with the standard unified model for AGN. This
should be investigated in more detail in the future, especially by studying
the luminosities
and the morphologies of the host galaxies in more detail.

\subsection{The distant dependent activity degree of AGN/non-AGN
galaxies}

We demonstrated that the [OIII] and H$\alpha$ line
luminosities in the environmental galaxies 
are increasing towards the central Seyfert and HII galaxies (Figs. 3 to 5).
These distant dependent line-intensity variations around Seyfert and HII
galaxies
are effective only for the emission line luminosities - not for the continuum
luminosities of the galaxies (Fig. 6).
The Seyfert~2 galaxies are on average slightly brighter
in the continuum than HII
galaxies (see Fig. 6). That trend in the continuum luminosities has been noted
before for nearby galaxy by Ho et at. (\cite{ho97}): They also
find that AGN are slightly brighter
than HII galaxies in the blue continuum (their Fig. 3). 

So far no thorough statistical analysis has been carried out
 of the emission line properties
in the large-scale environment around AGN/non-AGN.
Alonso et al. (\cite{alonso07}) performed a statistical analysis
of the [OIII] luminosities of AGN in
pairs and AGN without close companions. They find that the [OIII]
luminosity of AGN is correlated with the luminosity of the close companion
galaxies (r$_{proj}$ $\le$ 25 kpc).

However, we are investigating the large-scale environment of AGN and are
looking for the activity degree of the environmental galaxies. 
We demonstrated that
the emission line activity of the companions
is getting stronger up to a factor
of four within projected distances of 400 kpc around
central Seyfert and HII galaxies. 
The same trend toward increasing emission line luminosity in the
environmental galaxies
 as a function of projected proximity to the central AGN
has been noted before by
Kollatschny \& Fricke (\cite{kollatschny89}). This earlier study is
based on a much smaller sample of only nearby Seyfert galaxies.

The observed distant dependent activity degree can be explained by mutual
tidal triggering of Seyfert and starburst/HII activity.
If the duration of the starburst is around
$2\cdot{}10^8$ years (e.g. McQuinn et al., \cite{mcquinn09} and references
therein)
and if the internal velocity dispersion in groups/clusters is around
 1,000 $km~s^{-1}$ (e.g. Barrena et al., \cite{barrena11}; 
Fadda et al., \cite{fadda96}), then both interacting partners
 might have moved away by 200 kpc on average.
This distance covered within the lifetime of the starburst activity
corresponds to the results seen in Fig.~5. 
It indicates that both the starburst and the Seyfert activity are triggered
by tidal interaction and that the lifetimes of starburst and Seyfert
activities are of the same order of magnitude.
%

\begin{acknowledgements}
This work is based on SDSS data.
Funding for the creation and distribution of the SDSS Archive has been provided by the Alfred P. Sloan Foundation, the Participating Institutions, the National Aeronautics and Space Administration, the National Science Foundation, the U.S. Department of Energy, the Japanese Monbukagakusho, and the Max Planck Society. The SDSS Web site is http://www.sdss.org/.
The SDSS is managed by the Astrophysical Research Consortium (ARC) for the Participating Institutions. The Participating Institutions are The University of Chicago, Fermilab, the Institute for Advanced Study, the Japan Participation Group, The Johns Hopkins University, the Korean Scientist Group, Los Alamos National Laboratory, the Max-Planck-Institute for Astronomy (MPIA), the Max-Planck-Institute for Astrophysics (MPA), New Mexico State University, University of Pittsburgh, University of Portsmouth, Princeton University, the United States Naval Observatory, and the University of Washington.

 This work has been supported by the
Niedersachsen-Israel Research Cooperation Programm ZN2318 and
DFG grant Ko 857/32-1.
\end{acknowledgements}

{\it Note added in proof.}
In a recent publication, Haas et al. (\cite{haas12}) demonstrate a strong
correlation between the number of galaxy neighbors and the host's dark matter
halo. Combining their finding with our result indicates an inverse
dependence of star formation and/or AGN activity on the host's halo mass.

\newpage
\end{document}